\documentstyle[prl,aps,preprint,epsfig]{revtex}
\tightenlines
\newcommand{\be}{\begin{equation}}
\newcommand{\ee}{\end{equation}}
\newcommand{\ba}{\begin{eqnarray}}
\newcommand{\ea}{\end{eqnarray}}

\begin{document}
\draft

\title{Quantum  entropy of the Kerr
black hole\\ arising from gravitational perturbation}
  \author{Jiliang Jing $^{* \ a\ b }$\footnotetext[1]
 {email: jljing@hunnu.edu.cn} \ \
 \ \ Mu-Lin Yan $^{\dag\ b}$\footnotetext[2]
 {email: mlyan@ustc.edu.cn}}
\address{a) Institute of Physics and Physics Department, Hunan Normal
University,\\ Changsha, Hunan 410081, P. R. China;  \\ b)
Department of Astronomy and Applied Physics, University of Science
and Technology of China, \\ Hefei, Anhui 230026, P. R. China}
\maketitle
\begin{abstract}

The quantum entropy of the Kerr black hole arising from
gravitational perturbation is investigated by using Null tetrad
and 't Hooft's brick-wall model. It is shown that effect of the
graviton's spins on the subleading correction is dependent of the
square of the spins and the angular momentum per unit mass of the
black hole, and contribution of the logarithmic term to the
entropy will be positive, zero, and negative for different value
of $a/r_+$.

\end{abstract}
\vspace*{0.5cm}

By comparing black hole physics with thermodynamics and from the
discovery of black hole evaporation,  Bekenstein and Hawking
\cite{Bekenstein72} \cite{Hawking75} found that black hole
entropy is proportional to area of the event horizon, i.e.,
$S=A_H/4$. The discovery is one of the most profound one in black
hole physics. However, the issue of the exact statistical origin
of the black hole entropy, i.e., what degrees of freedom are
counted by the entropy of black holes, has remained a challenging
one. Recently, much effort has been concentrated on the problem
\cite{Hooft85} -\cite{Mann96}. 't Hooft \cite{Hooft85} proposed a
``brick wall" model (BWM) in which the black hole entropy is
identified with the statistical-mechanical entropy arising from a
thermal bath of quantum fields propagating outside the horizon.
The BWM was used to the studies of the statistical-mechanical
entropy arising from scalar fields for static black holes
\cite{Hooft85}  \cite{Jing98}\cite{Ghosh94} and for stationary
axisymmetric black holes \cite{Jing99}. The method also be
applied to calculate quantum entropy due to the electromagnetic
field for the Reissner-Nordstr\"{o}m black hole\cite{Cognola98}
and for the general static spherical static black
holes\cite{Jing2000I}. Recently  by using the BWM we
\cite{Jing2001} investigated the effects of spins of the photons
and Dirac particles on the entropies of the Kerr-Newman black
hole.

It has been believed that a black hole can exist in thermal
equilibrium with a heat bath  possessing a characteristic
temperature distribution. The heat bath could cause the change of
the space-time geometry by back reaction. From the semiclassical
Einstein equations $R_{\mu \nu}-\frac{1}{2}g_{\mu \nu }R= \langle
T_{\mu \nu}\rangle$ we know that  general metric can be
approximately written as $
 g_{\mu \nu}= g^A_{\mu \nu}+ g^B_{\mu \nu},
\label{gg} $
 where $ g^A_{\mu \nu}$ represents the classical
background space-time and $g^B_{\mu \nu}$ the disturbance. An
interesting open question is how the gravitational perturbation
affects the entropy of the black holes. The purpose of this paper
is to investigate the question by a concrete example, i.e., study
the quantum entropy of the Kerr black hole \cite{Kerr63} arising
from the gravitational disturbance by using the BWM.

A straightforward way to get perturbation equations for
gravitation is to insert the general metric given above into the
Einstein field equations and then obtain linear equations for the
perturbation. But even in the simplest static and spherically
symmetric case, to decouple the perturbation equations involves
considerable algebraic complexity. Therefore, in the case where
the background metric is stationary, the replacement of spherical
symmetry by axial symmetry means that to decouple the equations
is no longer possible by this way. Fortunately, there is an
alternative approach to the problem, which is provided by the
null tetrad formalism. In the following, we first introduce the
null tetrad to decouple gravitational perturbation equations,
then we seek the total number of modes under proper gauge, and
after that we calculate a free energy and the quantum entropy of
the Kerr black hole.

To decouple gravitational perturbation equations  in spacetime of
the Kerr black hole in the Boyer-Lindquist coordinates $(t,\ r,\
\theta,\ \varphi)$, we introduce null tetrad
 \ba
l^A_\mu&=&\frac{1}{ \bigtriangleup}( \bigtriangleup,\ \ -\Sigma ,\
\ 0,\ \ \ -a \bigtriangleup sin^2\theta ), \nonumber \\
 n^A_\mu&=&\frac{1}{
2\Sigma}( \bigtriangleup,\ \ \Sigma,\ \ 0,\ \ \ -a \bigtriangleup
sin^2\theta ), \nonumber \\
 m^A_\mu&=&-\frac{\bar{\rho}}{\sqrt{2}}(i a \sin\theta,
  \ \ 0, \ \ -\Sigma, \ \
-i(r^2+a^2)\sin\theta),\nonumber
\\
 \bar{m}^A_\mu &=&-\frac{\rho}{\sqrt{2}}(-i a \sin\theta,
 \ \ 0, \ \ -\Sigma, \ \ i(r^2+a^2)\sin\theta), \label{tetrad}
 \ea
here and hereafter the superscripts ``A" represent the undisturbed
values in the stationary spacetime, $\rho=-\frac{1}{r-i a
\cos\theta}$, $ \Sigma=r^2+a^2cos^2\theta, $
$\bigtriangleup=(r-r_+)(r-r_-), $ $r_{\pm}=M\pm\sqrt{M^2-a^2}$,
and $r_+$, $M$, and $a$ represent the radius of the event
horizon, the mass, and the angular momentum per unit mass of the
Kerr black hole, respectively.  The nonvanishing unperturbed
spin-coefficients and components of the Weyl tensor of the Kerr
black hole in the null tetrad (\ref{tetrad}) can then be expressed
 as\cite{Chandrasekhar92}\cite{Carmeli82}
 \ba
\rho^A&=&-\frac{1}{r-i a \cos\theta}, \ \ \ \ \
\beta^A=-\frac{\bar{\rho}cot\theta}{2\sqrt{2}}, \ \ \ \
\pi^A=\frac{i a \rho^2\sin\theta}{\sqrt{2}}, \ \ \ \
\tau^A=-\frac{i a \rho\bar{\rho}\sin\theta}{\sqrt{2}},\nonumber \\
\ \ \ \ \mu^A &=&\frac{ \rho^2\bar{\rho}\bigtriangleup}{2}, \ \ \
\gamma^A=\mu^A+\frac{\rho\bar{\rho}(r-M)}{2}, \ \ \ \
\alpha^A=\pi^A-\bar{\beta^A}, \ \ \ \ \psi^A_2=M\rho^3.
 \ea
When the stationary Kerr black hole is gravitational perturbed by
the incidence of gravitational waves, the quantities which vanish
in the stationary state will become quantities of the first order
of smallness and can be described by $ \psi_0, $ $ \psi_1, $
$\psi_3,$ $\psi_4, $ $ \kappa,$ $ \sigma,$ $ \lambda, $   and  $
\nu.$ We know from Ref. \cite{Chandrasekhar92} that, in a linear
perturbation theory, $\psi_0$ and $\psi_4$ are gauge invariant
quantities while $\psi_1$ and $\psi_3$ are not. Consequently, we
may choice a gauge (i.e., subject the tetrad basis to an
infinitesimal rotation) in which $ \psi_1$ and $ \psi_3 $ vanish
without affecting $\psi_0$ and $\psi_4$. If we choice such a
gauge and assume that
 \ba
\psi_0&=&R_{+2}(r)\Theta_{+2}(\theta)e^{-i(E t-m\varphi)},
\nonumber \\ \psi_4&=&\frac{1}{\bar{\rho} ^4}
R_{-2}(r)\Theta_{-2}(\theta)e^{-i(E t-m\varphi)}, \ \ \ \ ({\rm \
{\sl E}\  and\  {\sl m} \ are\  constants}) \label{psiij}
 \ea
we find that $\kappa$, $\sigma$, $\lambda$, and $\nu$ can be
written as \cite{Chandrasekhar92}
 $
 \kappa=-\frac{\sqrt{2}}{6M}\bar{\rho}^2 R_{+2}\left(
 {\cal{L}}_2-\frac{3 i a\sin \theta}{\bar{\rho}}
 \right)\Theta_{+2}, $ $
 \sigma=\frac{1}{6M}\frac{\bar{\rho}^2}{\rho}
 \Theta_{+2}\bigtriangleup
 \left( {\cal{D}}^{\dag}_2-\frac{3}{\bar{\rho}}\right)R_{+2},
 $$ \nu=\frac{\sqrt{2}}{6M}\frac{1}{\rho ^2}
 R_{-2}\left(
 {\cal{L}}^{\dag}_2-\frac{3 i a\sin \theta}{\bar{\rho}}
 \right)\Theta_{-2},$ $
\lambda=\frac{1}{6M}\frac{2}{\bar{\rho}}
 \Theta_{-2} \left(
 {\cal{D}}_0-\frac{3}{\bar{\rho}}\right)R_{-2},
  $
and decoupled equations are given by \cite{Chandrasekhar92}
 \ba \label{decoupled}
&&  (\bigtriangleup {\cal{D}}_1 {\cal{D}}^{\dag}_2-6iEr)
 R_{+2}(r)=\lambda_{+2} R_{+2}(r),
 \nonumber \\
&&   (\bigtriangleup {\cal{D}}^{\dag}_{-1} {\cal{D}}_0+6iEr)
 R_{-2}(r)=\lambda_{-2} R_{-2}(r),
 \nonumber \\
 &&
({\cal{L}}^{\dag}_{-1}{\cal{L}}_{2}+6aE\cos \theta )
 \Theta_{+2}(\theta)=-\lambda_{+2} \Theta_{+2}
 (\theta),\nonumber \\
&&({\cal{L}}_{-1}{\cal{L}}^{\dag}_{2}-6aE\cos \theta )
 \Theta_{-2}(\theta)=-\lambda_{-2} \Theta_{-2}
 (\theta),
 \ea
where $\lambda_{+2}$ and $\lambda_{-2}$ are separation constants,
and
 $
 {\cal{D}}_n\equiv \frac{\partial}{\partial r}
 +\frac{i (r^2+a^2)E-ma}
 {\bigtriangleup}+2n\frac{r-M}{\bigtriangleup},$ \
${\cal{D}}^{\dag}_n\equiv\frac{\partial}
 {\partial r}-\frac{i (r^2+a^2)E-ma}
 {\bigtriangleup}+2n\frac{r-M}{\bigtriangleup},$\
$ {\cal{L}}_n\equiv\frac{\partial}{\partial \theta}
 +a E \sin \theta-\frac{m}{\sin
\theta }
 +n\cot \theta,$\  and
 ${\cal{L}}^{\dag}_n\equiv\frac{\partial}{\partial \theta}-a E \sin
\theta-\frac{m}{\sin \theta }
 +n\cot \theta. \label{ld}$
 The Eq. (\ref{decoupled}) can be explicitly expressed as
 \ba
&& \bigtriangleup \frac{d^2R_{s}}{d r^2}+6(r-M)
 \frac{dR_{s}}{d r}+\left[
 2s+4i s r E+\frac{K_1^2-2i s K_1 (r-M)}{ \bigtriangleup}
 -\lambda_{s}^2\right]R_{s}
 =0,  \ \ \ \   (s=+2),\nonumber \\
&& \bigtriangleup\frac{d^2R_{s}}{d r^2}-2(r-M)
 \frac{d R_{s}}{d r}+\left[
 +4 i s r E+\frac{K_1^2-2i s K_1 (r-M)}
 {\bigtriangleup}-\lambda_{s}^2\right]R_{s}
 =0,  \ \ \ \ \  \ \ \   (s=-2), \nonumber \\
&& \frac{d^2\Theta_{s}}{d \theta^2}+\cot \theta
 \frac{d\Theta_{s}}{d \theta}+\left[2m a E
 -a^2E^2\sin^2\theta-\frac{m^2}{\sin^2\theta}\right. \nonumber \\
&&\hspace*{4.0cm}\left. +2a s E \cos \theta
 +\frac{2 s m \cos \theta}{\sin^2\theta}-s-s^2\cot
 ^2\theta+\lambda_{s}^2\right]\Theta_{s}
 =0,   \ \ \ \ \   (s=+2),\nonumber \\
&& \frac{d^2\Theta_{s}}{d \theta^2}+\cot \theta
 \frac{d\Theta_{s}}{d \theta}+\left[2m a E
 -a^2E^2\sin^2\theta-\frac{m^2}{\sin^2\theta}\right. \nonumber \\
&&\hspace*{4.0cm}\left. +2a s E \cos \theta
 +\frac{2 s m \cos \theta}{\sin^2\theta}+s-s^2\cot
 ^2\theta+\lambda_{s}^2\right]\Theta_{s}=0,   \ \ \ \ \
  (s=-2). \nonumber \\ \label{rfrf}
 \ea
where $K_1=(r^2+a^2)E-m a$.
 We now adopt WKB approximation by writing
the mode functions as
 $
R_{s}(r)=\tilde{R}_{s}(r) e^{-i k_{s}(E,m,k_s(\theta),r, \theta)
r},
 $ $
 \Theta_{s}(\theta)=\tilde{\Theta}
 _{s}(\theta)e^{-i k_{s}(\theta) \theta},
 $
and supposing that the amplitudes $ \tilde{R}_{s}(r)$ and $
\tilde{\Theta}_{s}(\theta)$ are slowly varying functions, that is
to say,
 $
 \left|\frac{1}{\tilde{R}_{s}}
 \frac{d\tilde{R}_{s}}{d r}\right| \ll |k_{s}(E,m,k_s(\theta),r,\theta)|,$ $
 \left|\frac{1}{\tilde{R}_{s}}
 \frac{d^2\tilde{R}_{s}}{d r^2}\right| \ll |k_{s}(E,m,k_s(\theta),r,\theta)|^2, $
 $
  \left|\frac{1}{\tilde{\Theta}_{s}}
 \frac{d\tilde{\Theta}_{s}}{d r}\right| \ll |k_{s}(\theta)|,
$ and $
 \left|\frac{1}{\tilde{\Theta}_{s}}
 \frac{d^2\tilde{\Theta}_{s}}{d r^2}\right|
 \ll |k_{s}(\theta)|^2.
 $
Thus, from Eqs. (\ref{psiij}) and (\ref{rfrf}) we know that both
$k_{+2}(E,m,k_{+2}(\theta),r,\theta)$ for $\psi_0$ and
$k_{-2}(E,m,k_{-2}(\theta),r,\theta)$ for $\psi_4$ can be
expressed as
 \ba
 &&k_{s}(E,m,k_s(\theta),r,\theta)
 ^2=\frac{\left[(r^2+a^2)E-ma\right]^2}
 { \bigtriangleup^2}+\frac{1}{ \bigtriangleup} \left(2 m a E
 -a^2E^2\sin^2\theta-\frac{m^2}{\sin^2\theta }
\right. \nonumber \\ &&\left.\hspace*{4.2cm} -
 k_{s}({\theta})^2
 +2 s a E\cos\theta
 +\frac{2s m \cos\theta}{\sin^2\theta} +s-s^2cot^2\theta\right),
 \label{psi11}\nonumber \\
&&\hspace*{4.2cm}(s=+2\ \  {\rm for}\ \  \psi_0,\ \  {\rm and } \
\ s=-2 \ \ {\rm for}\ \  \psi_4).
 \ea
 The
number of modes for each component $\psi_{i}$ with $E$, $m$ and
$k_{\theta}$ takes the form
 $ n_{s}(E, m, k_{s}({\theta}))=\frac{1}{\pi}
\int d\theta \int^{L}_{r_++h} d r
k_{s}(E,m,k_s(\theta),r,\theta).\label{nijij}
 $
Here we introduced 't Hooft's brick-wall boundary conditions: the
gravitational filed wave functions are cut off outside the
horizon, i.e., $\psi_{0}=\psi_{4}=0$ at $\Sigma_h$ which stays at
a small distance $h$ from the event horizon $r_+$. There is also
an infrared cutoff $\psi_{0}=\psi_{4}=0$ at $r=L$, where the
infrared cutoff $L$ is chosen so that the quantum gas is inside
the null cylinder, a surface where the co-rotation velocity
reaches the velocity of light.

Thermal equilibrium between a quantum gas and a stationary
axisymmetric black hole at temperature $1/\beta$ is only possible
when the gas is rigidly rotating with the angular velocity equal
to the velocity of the black hole horizon $\Omega_H$. Therefore,
it is rational to assume that the gravitational field is rotating
with angular velocity $\Omega_0=\Omega_{H} $ near the event
horizon. For such an equilibrium ensemble of states, the free
energy is
\begin{eqnarray}
\beta F&=& \int d m \int d k_s({\theta})\int d n_s(E, m,
k_s({\theta}))\ln \left[ 1-e^{-\beta (E-\Omega_0 m)}\right]
\nonumber  \\ &=&-\beta \int d m \int d k_s({\theta})\int
\frac{n_s(E+\Omega_0 m, m, k_s({\theta}))} {e^{\beta E}-1} d E
\nonumber \\ &=&-\beta \int \frac{n(E)}{e^{\beta E}-1} d E ,
\label{f1}
\end{eqnarray}
with \ba n(E)\equiv\sum_s n_s(E)=\sum_s \int d m \int d
k_s({\theta})\int n_{s}(E+\Omega_0 m, m, k_s({\theta})),
\label{nsum} \ea where the function $n(E)$ presents the total
number of modes with energy less than $E$. In order to carry out
the calculation for the $n(E)$ we recast Eqs. (\ref{psi11}) into
the forms
 \ba
&& k_{s}(E,m,k_s(\theta),r,\theta)=\sqrt{\frac{-g_{rr}
g_{\varphi\varphi}}{g_{tt}g_{\varphi\varphi}-
g_{t\varphi}^2}}\left\{(E-m\Omega)^2+\left(g_{tt}-
\frac{g_{t\varphi}^2} {g_{\varphi\varphi}}\right)\left[
\frac{k_s(\theta)^2}{g_{\theta\theta}}+\left(\frac{m}
{\sqrt{g_{\varphi\varphi}}}\right.\right.\right. \nonumber
\\ && \left. \left. \left. -\frac{s\sqrt{g_{\varphi\varphi}}
\cos\theta}{g_{\theta\theta}\sin^2\theta}\right)^2
+\frac{s^2}{g_{\theta\theta}}
\left(1-\frac{g_{\varphi\varphi}}{g_{\theta\theta}\sin^2\theta}\right)
\cot^2\theta-\frac{s}{g_{\theta\theta}}(1-2aE\cos\theta)
\right]\right\}^{1/2}, \hspace*{0.5cm}(s=\pm2),\label{k11}
 \ea
where the function $\Omega\equiv-\frac{g_{t\varphi}}{g_{\varphi
\varphi}}$ and its value on the event horizon is equal to
$\Omega_H$. Substituting Eq. (\ref{k11}) into (\ref{nsum}) and
carrying out the integrations we find
\begin{eqnarray}
n(E)&=&\frac{4E^3}{3\pi}\left(\frac{\beta_H}{4\pi}\right)^3\int
d\theta \left[\sqrt{g_{\theta\theta}g_{\varphi\varphi}}\left(
\frac{1}{h}\frac{\partial g^{rr}}{\partial r}-
\left\{\frac{\partial ^2g^{rr}}{\partial
r^2}+\frac{3}{2}\frac{\partial g^{rr}}{\partial r}\frac{\partial
\ln f}{\partial
r}\right.\right.\right. \nonumber \\
&&-\left.\left.\left.\frac{2\pi}{\beta_H\sqrt{f}}
\left(\frac{1}{g_{\theta \theta}}\frac{\partial g_{\theta
\theta}}{\partial r}+ \frac{1}{g_{\varphi \varphi}}\frac{\partial
g_{\varphi \varphi}}{\partial r}\right)-\frac{2g_{\varphi
\varphi}}{f}\left[\frac{\partial}{\partial r}\left(\frac{g_{t
\varphi}}{g_{\varphi \varphi}}\right)\right]^2\right\}
\ln\frac{L}{h} \right)\right]_{r_+} \nonumber \\
&&+\frac{2s^2E}{\pi}\left(\frac{\beta_H}{4\pi}\right)\int d\theta
\left[\sqrt{g_{\theta\theta}g_{\varphi\varphi}}\left(1-\frac{g_{\varphi\varphi}}
{g_{\theta\theta}\sin^2\theta}\right)\frac{\cot^2\theta}
{g_{\theta\theta}}\right]_{r_+}\ln\frac{L}{h}, \label{n0}
\end{eqnarray}
where  $f\equiv
-g_{rr}\left(g_{tt}-\frac{g_{t\varphi^2}}{g_{\varphi\varphi}}\right)$.
With the aid of the  expression  (\ref{n0}), we can work out the
free energy defined by Eq. (\ref{f1}). Then, the relation between
the entropy and the free energy, $ S=\beta^2\frac{\partial
F}{\partial \beta}$, shows that
\begin{eqnarray}
S&=&\frac{ A_{H}}{24\pi\epsilon^2}- \frac{1}{180}\int d\theta
\left\{\sqrt{g_{\theta \theta}g_{\varphi \varphi}}\left[
\frac{\partial ^2g^{rr}}{\partial r^2}+ \frac{3}{2}\frac{\partial
g^{rr}}{\partial r}\frac{\partial \ln f}{\partial
r}-\frac{2\pi}{\beta \sqrt{f}}\left(\frac{1} {g_{\theta
\theta}}\frac{\partial g_{\theta \theta}}{\partial r}+
\frac{1}{g_{\varphi \varphi}}\frac{\partial g_{\varphi
\varphi}}{\partial r}\right)  \right. \right.\nonumber \\ & -&
\left. \left.   \frac{2g_{\varphi \varphi}}{f}
\left(\frac{\partial}{\partial r}\frac{g_{t \varphi}}{g_{\varphi
\varphi}}\right)^2\right]+ \frac{s^2}{6}\int
d\theta\left[\sqrt{g_{\theta \theta}g_{\varphi
\varphi}}\left(1-\frac{g_{\varphi\varphi}}
{g_{\theta\theta}\sin^2\theta}\right)\frac{\cot^2\theta}
{g_{\theta\theta}}\right]_{r_+}\right\}_{r_+}
\ln\frac{\Lambda}{\epsilon}, \label{smu}
\end{eqnarray}
where $\delta^2=\frac{2\epsilon^2}{15}$ and
$\Lambda^2=\frac{L\epsilon ^2}{h}$   \cite{Jing98} (where
$\delta=\int_{r_+}^{r_++h}\sqrt{g_{rr}}d r\approx
2\sqrt{h/\left(\frac{\partial g^{rr}}{\partial r}\right)_{r_+}}$
is the proper distance from the horizon to $\Sigma_h $,
$\epsilon$ is the ultraviolet cutoff, and $\Lambda$ is the
infrared cutoff \cite{Solodukhin96}) , and $A_H=\int d \varphi
\int d \theta (\sqrt{g_{\theta \theta}g_{\varphi
\varphi}})_{r_+}$ is area of the event horizon. By inserting the
metric of the Kerr black hole into Eq. (\ref{smu}) we final find
that the quantum entropy of the Kerr black hole due to the
gravitational perturbation is given by
 \ba
S&=&\frac{A_{H}}{24\pi\epsilon^2}+
\left\{\frac{2}{45}+\frac{s^2}{6}\left[1-\frac{r_+^2+a^2}{a
r_+}\arctan\left(\frac{a}{r_+}\right)\right]\right\}
\ln\frac{\Lambda}{\epsilon}. \label{kn2} \ea

\begin{figure}
\centerline{
        \psfig{figure=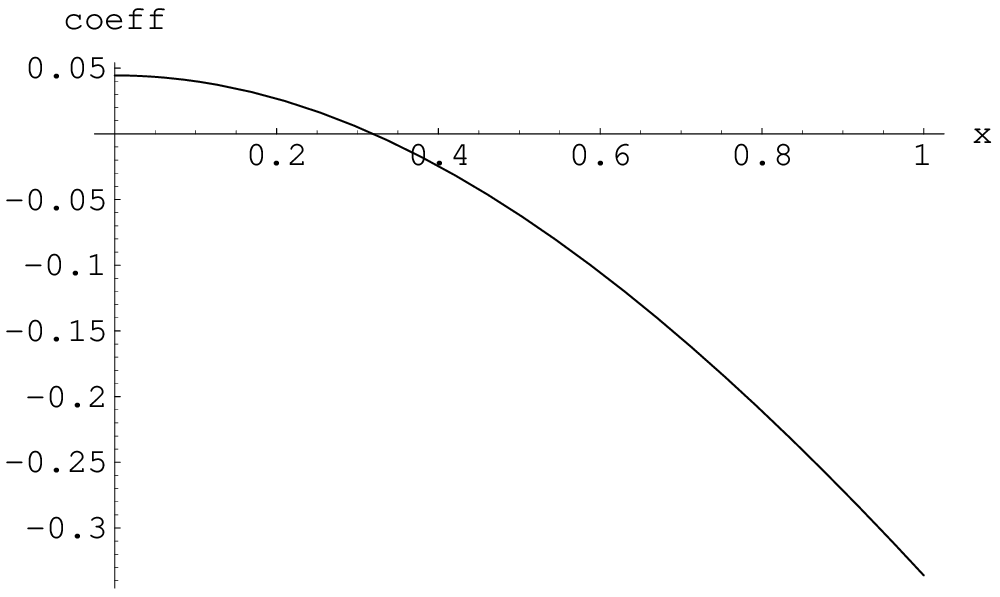,height=2.0in,angle=0}}
        \caption{The coefficient of the logarithm term runs from
        0.0444444 to -0.336086 as $x=\frac{a}{r_+}$ from 0 to 1,
        and is equal to zero at point $x\approx 0.319366$. It shows
        that the logarithmic term increases the entropy in range
$\frac{a}{r_+}= [0 ,0.319366)$;  decreases the entropy for
$\frac{a}{r_+}= (0.319366, 1]$; and does not affect the entropy
when $\frac{a}{r_+}\approx 0.319366$. The reason is that the
terms for $s^2$ decrease the entropy except static case, $a=0$.
        }
\label{fig1}

\end{figure}

Several remarks regarding main result (\ref{kn2}) of the paper
are in order: (I)\ The result is different from quantum entropy
of the Kerr black hole caused by the scalar field which coincides
with that of the Schwarzschild black hole\cite{Mann96}, i.e.,
$S_{Scalar, Kerr}=\frac{A_{H}}{48\pi\epsilon^2}+ \frac{1}{45}
\ln\frac{\Lambda}{\epsilon}$. The disagreement exists even if s=0
in Eq. (\ref{kn2}) (by an overall 2 factor). The discrepancy is
originated by summing over polarizations. (II)\ The Fig.
\ref{fig1} shows that the logarithmic term will increase the
entropy in range $\frac{a}{r_+}= [0 ,0.319366)$; and decrease the
entropy for $\frac{a}{r_+}= (0.319366, 1]$. The term does not
affect the entropy when $\frac{a}{r_+}\approx 0.319366$. The
reason is that the terms for $s^2$ decrease the entropy except
static case, $a=0$. (III) \ {\sl The subleading logarithmic
correction of the quantum entropy depends on the spins of the
graviton just in quadratic term $s^2$.} We know from each
component of the field that the number of modes for every
component field contains both terms of the $s$ and $s^2$.
However, the linear terms of $s$ are eliminated each other when
we sum over all components to get the total number of modes.
(IV)\ \ {\sl The contribution of the spins to the logarithmic
term is related to the rotation of the black hole.} For the
static case, i.e., $a=0$, the terms for $s^2$  in the results
(\ref{kn2}) vanishes. It is shows that the spins of the particles
affect the logarithmic correction of the entropy in case of
interaction between the spins of the particles and the rotation
of the black hole takes place. It should be noted that since the
$\psi_i$ can not be decoupled in the Kerr-Newman spacetime
\cite{Chandrasekhar92} the quantum entropy of the Kerr-Newman
black hole is an interesting problem and is being under
considered.

\begin{acknowledgements}
This work was supported by the National Natural Science Foundation
of China under Grant No. 19975018,  Theoretical Physics Special
Foundation of China under Grant No. 19947004, and National
Foundation of China through C. N. Yang.
\end{acknowledgements}

\end{document}